\documentclass[pra,showpacs,twocolumn]{revtex4-1}

\usepackage{amsmath}
\usepackage{graphicx}
\usepackage{bm}

\newcommand{\pa}{\partial}
\newcommand{\psiup}{\psi_\uparrow(x)}
\newcommand{\psidn}{\psi_\downarrow(x)}
\newcommand{\psiupdg}{\psi_\uparrow^\dagger(x)}
\newcommand{\psidndg}{\psi_\downarrow^\dagger(x)}
\newcommand{\hbarm}{\frac{\hbar^2}{2m}}

\begin{document}

\title{One-dimensional Fermi polaron in a combined harmonic and periodic potential}
\author{E. V. H. Doggen, A. Korolyuk, P. T\"orm\"a and J. J. Kinnunen}
\affiliation{COMP Centre of Excellence and Department of Applied Physics, Aalto University, FI-00076 Aalto, Finland}
\begin{abstract}
We study an impurity in a one-dimensional potential consisting of a harmonic and a periodic part using both the time-evolving block decimation (TEBD) algorithm and a variational ansatz. 
Attractive and repulsive contact interactions with a sea of fermions are considered. 
We find excellent agreement between TEBD and variational results and use the variational ansatz to investigate higher lattice bands.
We conclude that the lowest band approximation fails at sufficiently strong interactions and develop a new method for computing the Tan contact parameter.
\end{abstract}

\maketitle

\section{Introduction}

Impurities in lattices are of interest because of intriguing phenomena such as the Kondo effect \cite{Kondo1964a}, Anderson localization \cite{Anderson1958a} and colossal magnetoresistance \cite{Mannella2005a}.
One-dimensional (1D) systems in particular are appealing because analytical results are available for the homogeneous fermionic impurity interacting through a delta function potential \cite{McGuire1965a,McGuire1966a}.
Various schemes such as the T-matrix approach, time-evolving block decimation (TEBD) \cite{Vidal2003a}, quantum Monte Carlo simulations and a variational ansatz \cite{Chevy2006a} have been successfully applied to the problem of an impurity interacting with fermions in higher dimensions \cite{Lobo2006a, Combescot2007a, Combescot2008a, Massignan2011a, Parish2013a} as well as one dimension \cite{Giraud2009a, HeidrichMeisner2010a, Punk2009a, Astrakharchik2013a, Massel2013a, Mathy2012a, Doggen2013a} (for recent review articles, see Refs.\ \cite{Guan2013a,Massignan2014a}).
Interest in 1D systems has further increased after the realization of such systems in ultracold gases using optical lattices, for instance the Tonks-Girardeau gas \cite{Kinoshita2004a,Paredes2004a}.
Experimentally, the reduction in dimensionality is achieved by tightly confining the gas in two of the three spatial dimensions.
Recent experiments study impurities in 1D bosonic systems \cite{Catani2012a,Fukuhara2013a} and the 1D Fermi polaron in a few-body system \cite{Wenz2013a}.
Most theoretical studies of ultracold fermions in a lattice focus on the effects of the lattice, and for conceptual and numerical simplicity neglect the effect of the harmonic trap.
However, the inclusion of the harmonic trap changes the density of states \cite{Hooley2004a} and in experimental practice a trapping potential is always present to confine the cloud of atoms.

Recently it was proposed by Tan \cite{Tan2008b} that the high-momentum occupation probability $n_q$ of fermions interacting through a short-range potential obeys the universal relation $n_q \sim C/q^4$, where $q$ is momentum.
The quantity $C$ is called the \textit{contact}, because it is a measure of the probability of finding two particles in close proximity.
Remarkably, the contact contains all the information about the many-body properties of the system and is furthermore appealing because of the relative ease with which it is measured, for example using rf spectroscopy \cite{Stewart2010a}.
The Tan contact parameter was the subject of subsequent theoretical and experimental research (see Ref.\ \cite{Zwerger2012a} and references therein), although the majority of theoretical research was on homogeneous, spin-balanced systems (trapped systems have been discussed e.g.\ in Refs.\ \cite{Tan2011a,Yin2013a}).
In this work, we look at the strongly spin-imbalanced, non-homogeneous case.

We present a comprehensive study of an impurity in a 1D lattice with a harmonic trapping potential, interacting with a bath of majority component fermions.
We note, however, that our variational model imposes no \textit{a priori} restrictions on the external potential or the dimensionality of the system.
This paper is structured as follows. 
First, we investigate the problem in the lowest band approximation using both a variational ansatz and TEBD.
Secondly, we consider higher lattice bands and evaluate the validity of the lowest band approximation.
Finally, we discuss the high-energy excitations of the impurity and the associated contact parameter.

\section{The One-dimensional Fermi Polaron}

We consider an impurity (a ``spin down'' atom) immersed in a sea of $N$ (``spin up'') fermions, with an external potential consisting of a harmonic and a periodic part. 
This system is described in one dimension by the following Hamiltonian:
\begin{align}
   \mathcal{H} = & \sum_{\sigma}  \int dx  \, \psi_{\sigma}^\dagger(x)\Big[ -\frac{\hbar^2}{2m_\sigma} \frac{d^2}{dx^2} + V(x) \Big] \psi_\sigma(x) \nonumber \\
  &+ g \int dx \, \psiupdg \psidndg \psidn \psiup = \mathcal{H}_0 + \mathcal{H}_\text{int} \label{mainhamiltonian}
\end{align}
where $V(x) = \nu x^2 + \frac{V_0}{2} (1 - \cos(2\pi k_\text{L} x))$, $x$ is position, $m_\sigma$ is the mass of a particle, $\nu$ is the strength of the harmonic potential, $V_0$ is the depth of the lattice, $k_\text{L}$ determines the periodicity of the lattice, $g$ determines the strength of the inter-particle interaction (we assume contact interactions) and $\psi_\sigma^{(\dagger)}$ destroys (creates) a particle in the spin state $\sigma \in \{\uparrow,\downarrow \}$.
In practice, the ``spin'' states may, for instance, be different hyperfine states of the same atom.

We study this Hamiltonian using two distinct approaches: the TEBD algorithm \cite{Vidal2003a} and a different implementation of a variational ansatz proposed by Chevy \cite{Chevy2006a}.
The TEBD simulation employs the Hubbard Hamiltonian (with additional harmonic trapping):
\begin{align}
 \mathcal{H}_\text{Hubbard} = & - \sum_{i\sigma} J_\sigma c_{i\sigma}^\dagger c_{i+1\sigma} + h.c. + U \sum_i c_{i\uparrow}^\dagger c_{i\uparrow} c_{i \downarrow}^\dagger c_{i \downarrow} \nonumber \\
 & + V_\text{h} \sum_{i\sigma} c_{i\sigma}^\dagger c_{i\sigma} i^2, \label{hubbardhamiltonian}
\end{align}
where $J_\sigma$ is the hopping parameter (which can depend on spin), $U$ determines the strength of the inter-particle interaction, $c_{i\sigma}^{(\dagger)}$ destroys (creates) a particle with spin $\sigma$ at site $i$ (we choose $i = 0$ as the center site) and $V_\text{h}$ measures the strength of the harmonic trap.
The variational ansatz, on the other hand, is an approximative scheme where one restricts the Hilbert space to include at most a single particle-hole pair (while it is possible to generalize the ansatz to a higher number of particle-hole pairs \cite{Giraud2009a}, we will consider at most a single pair).
We use a more general form of the variational ansatz of Ref.\ \cite{Chevy2006a}, where instead of a polaron at fixed momentum we consider a general superposition (similar extensions were derived in Refs.\ \cite{Ku2009a, Levinsen2012a}):
 \begin{equation}
 |\Psi \rangle = \sum_l \phi_l a_{\downarrow l}^\dagger  |0\rangle + \sum_{mkn} \phi_{mkn} a_{\uparrow m}^\dagger a_{\uparrow k} a_{\downarrow n}^\dagger |0 \rangle. \label{RSVA}
\end{equation}
Here $\phi_l$ and $\phi_{mkn}$ ($m \neq k$) are variational parameters, which are to be determined, $a_{\sigma m}^{(\dagger)}$ destroys (creates) a particle with spin $\sigma$ in an occupied (empty) state $m$, $m = 0$ denotes the ground state of the non-interacting system and $|0\rangle$ is a shorthand notation for the ground state of the non-interacting system of $N$ fermions.
The operators $a_{\uparrow}^{(\dagger)}$ destroy (create) particles in the eigenstates of $\mathcal{H}_0$.
On the other hand, the operators $a_{\downarrow}^{(\dagger)}$ for the minority component atom correspond to the eigenstates of the ``mean-field'' Hamiltonian $\mathcal{H}_0 + gn_\uparrow(x)$ ($n_\uparrow(x)$ is the density of the majority component atoms, where the density distribution of the non-interacting gas is used), of which the ground state is expected to be closer to the ground state of the full Hamiltonian (\ref{mainhamiltonian}).
The computation then requires calculating the expectation value of the Hamiltonian $\langle \Psi| \mathcal{H} |\Psi \rangle$ and minimizing with respect to the variational parameters $\phi_l$ and $\phi_{mkn}$ using an iterative procedure (a detailed description is shown in the appendix).
In principle, one can do this in any basis for the minority and majority component atoms -- our choice is simply a matter of computational convenience.
We will consider eigenstates in real space only and henceforth refer to this method as the real-space variational ansatz (RSVA).

The Hubbard Hamiltonian (\ref{hubbardhamiltonian}) is known to describe physics limited to the first band of the lattice accurately.
TEBD gives essentially exact numerical results for the Hubbard model (\ref{hubbardhamiltonian}) and does not restrict the number of particle-hole excitations.
However, the Hubbard model does not reproduce all the features of the Hamiltonian (\ref{mainhamiltonian}); only the lattice sites are considered, and the effect of the lattice in the tight-binding approximation is accounted for through the simplified hopping parameter $J$.
While the RSVA gives only approximate results, it is easily extended beyond these limitations to enhance the spatial resolution and access higher bands.
The method is also numerically efficient; our implementation of the RSVA is three orders of magnitude faster than our TEBD implementation.

To provide the mapping between the full Hamiltonian (\ref{mainhamiltonian}) and the Hubbard Hamiltonian (\ref{hubbardhamiltonian}), we express energies in terms of the recoil energy $E_\text{R} = \frac{\hbar^2 k_\text{L}^2}{2m}$ and scale lengths by $k_\text{L}$.
We will first consider the case where the full Hamiltonian (\ref{mainhamiltonian}) is discretized in space with spacing $1/k_\text{L}$, such that only the bottoms of the lattice wells are considered and the parameter $V_0$ plays no role.
This corresponds to the lowest band approximation (LBA) of the single-band Hubbard Hamiltonian.
We use closed boundary conditions throughout this paper.

\section{Results}

\begin{figure}[!htb]
 \centering
 \includegraphics[width=\columnwidth]{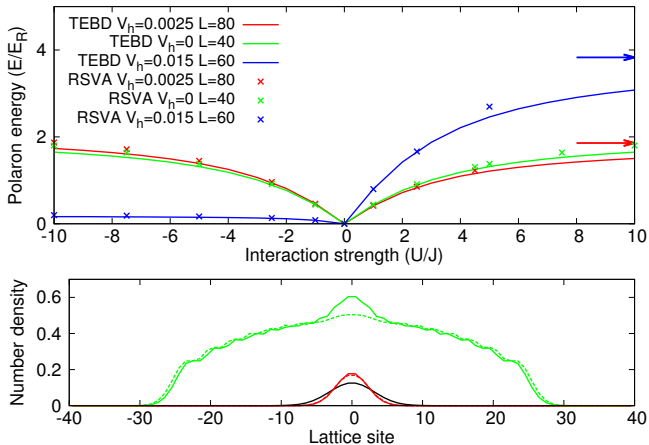}
 \caption{(color online). Top: polaron energy $E_\text{pol}$ as a function of the interaction strength $U/J$ for a system with $N = 20$, various harmonic trapping strengths $V_\text{h}$ and number of lattice sites $L$. 
 Solid lines show the TEBD prediction, crosses show the RSVA prediction.
 Arrows show the $U \rightarrow \infty$-limit for $V_\text{h} = 0.0025$ and $0.015$, obtained by considering $N+1$ spinless fermions.
 For the variational ansatz, the iteration fails to converge at $U/J \approx 5$ for $V_h \neq 0$.
 On the attractive side ($U/J < 0$) the maximum on-site interaction energy $U/J$ has been subtracted.
 Bottom: number density profiles computed using the TEBD (solid lines) and RSVA (dashed lines) methods for $U/J = -10$ and $V_\text{h} = 0.0025$. 
 The green line shows the majority component density and the red line the polaron density. 
 For comparison, the solid black line shows the ground state of the polaron in the non-interacting system.} \label{tebdenergy}
\end{figure}

The polaron energy $E_\text{pol}$ is defined as the energy of the impurity minus the energy of the impurity in the non-interacting system.
In Fig.\ \ref{tebdenergy} we show $E_\text{pol}$ as a function of $U/J$ as computed by the RSVA and TEBD, for various trapping frequencies.
The agreement on the attractive ($U/J < 0$) side is excellent, while on the repulsive side we find good agreement for weak to moderate ($0 < U/J \lesssim 4$) interactions.
For example, for a harmonic trap frequency of $V_h/E_\text{R} = 0.0025$ we find a relative error of $1.7\%$ at $U/J = -5$ and an error of $0.2\%$ at $U/J = 1$.
In the case of a finite trap, the iteration fails to converge at stronger ($U/J \gtrsim 5$) repulsive interactions.
This is most likely due to the restrictions of the ansatz, which is not self-consistent in the majority component density.
The inset of Fig.\ \ref{tebdenergy} also shows that while the energy matches very well with exact results, the prediction for the density profiles shows only qualitative agreement (see also the discussion concerning quasiparticle weight in Ref.\ \cite{Punk2009a}).
Considering strongly repulsive interactions using the TEBD method we find, perhaps counter-intuitively, the impurity in the center of the trap.
Here it has a strong density overlap with the majority component.
This is possible because of an arrangement where the doublon density $\langle c_{i\downarrow}^\dagger c_{i\downarrow} c_{i\uparrow}^\dagger c_{i\uparrow} \rangle$ is zero; the ground state is then a superposition of states with single occupancy (either $\downarrow$ or $\uparrow$) of each lattice site.

The almost exact match on the attractive side is in agreement with results for the homogeneous case \cite{Giraud2009a}.
For strongly attractive interactions, the energy is given by $E/E_\text{R} = U/J + E_\text{offset}/E_\text{R}$, where $E_\text{offset}$ depends on the majority component number density $\bar{n}$ near the center.
This can be understood as follows.
As the interaction becomes strongly attractive, the impurity will effectively pair with one of the majority component atoms, thus resulting in an interaction energy $U/J$.
However, this requires a rearrangement of the particles, resulting in a kinetic and trap energy penalty which depends only on $V_\text{h}$ and $N$.
Indeed, we find that the energy of the trapped system with $V_\text{h}/E_\text{R} = 0.0025$ and number of lattice sites $L = 80$ is almost the same as the untrapped ``lattice in a box'' with $L = 40$.
In both cases, $\bar{n} \approx 0.5$.
This is consistent with the finding that a theory based on the local density approximation works well \cite{Astrakharchik2013a}.
In the tightly trapped limit $V_h \rightarrow \infty$ particles are forced into the center of the trap, yielding unit filling $\bar{n} = 1$ for the $N$ sites near the center, and we recover $E_\text{pol} = U/J$.
Note that the special case of a harmonic trap with $N = 1$ was solved analytically \cite{Busch1998a}.

\begin{figure}[!htb]
 \centering
 \includegraphics[width=\columnwidth]{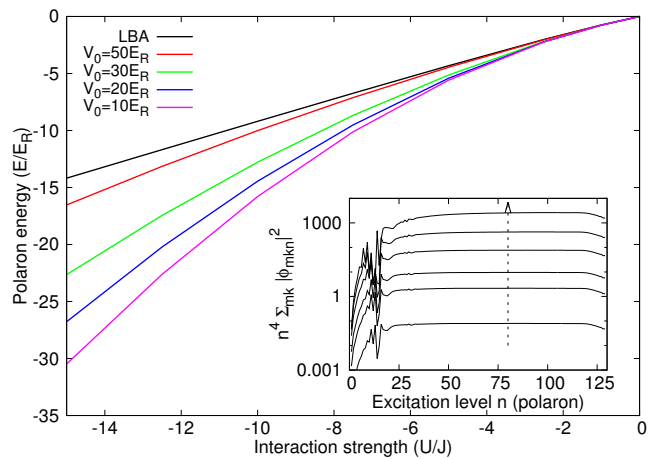}
 \caption{(color online). Polaron energy $E_\text{pol}$ as a function of the interaction strength $U/J$ for a system with $V_\text{h}/E_\text{R} = 0.1$, $N = 6$, $L = 16$ and the first 8 Bloch bands, for various values of the lattice depth $V_0$. The black solid line shows the LBA. All values calculated using eq.\ (\ref{RSVA}). Inset (note the log scale): occupation probability of excited states $\sum_{mk} |\phi_{mkn}|^2$ multiplied by $n^4$, showing the characteristic decay. We have verified by considering more than 8 Bloch bands that the drop at high $n$ is due to cutoff effects. $-U/J = 0.1,0.5,1.0,2.5,5.0,10.0, V_0/E_\text{R} = 10$. The dashed arrow indicates increased values of $|U/J|$.} \label{resolved}
\end{figure}

With the RSVA in solid footing, we move to investigate the effect of higher lattice bands.
While the Hubbard Hamiltonian is restricted to lattice sites, the variational approach allows spatially resolving the lattice, or indeed any spatially dependent potential, easily.
This introduces a new free parameter $V_0/E_\text{R}$, which describes the depth of the lattice.
Fig.\ \ref{resolved} shows $E_\text{pol}$ as a function of $U/J$ for various values of $V_0/E_\text{R}$ using 6 majority component particles and a harmonic trap $V_\text{h}/E_\text{R} = 0.1$.
We resolve the lattice in real space with $L = 16$ sites using $128$ points per lattice site and restrict the calculation to the first 8 Bloch bands.
Convergence with the number of points per lattice site is fast and increasing the number further does not provide a significant change to the energy.
Conversely, the convergence of the energy with the number of Bloch bands is slower \cite{Buchler2010a}; including the ninth band gives a relative correction of $1.2\%$ at $U/J = -10$ and $V_0/E_\text{R} = 10$.
The proper rescaling of $U/J$ as a function of $V_0/E_\text{R}$ is obtained self-consistently by demanding that in the limit of small $U/J$ the LBA is accurate.
For high enough $V_0/E_\text{R}$ the prediction from the LBA is quite accurate, even for strongly attractive interactions, as expected.
However, as $V_0/E_\text{R}$ is reduced, its accuracy quickly deteriorates.
The effect of the harmonic trap (which varies significantly over lattice sites) increases the lattice depth required to be able to apply the LBA.
In the lattice-only case, the LBA is known to be accurate as long as $|U/J| \lesssim V_0/E_\text{R}$ (see \cite{bloch2008a} and references therein).
We have verified by comparing to the case $V_\text{h} = 0$ that for $U/J=-10$ and $V_0/E_\text{R} = 10$, about two thirds of the deviation from the LBA is due to harmonic confinement.

For any finite interaction strength, a fraction of the particles occupy all higher lattice bands.
This allows access to the occupation probability of highly excited states, from which one can obtain the Tan contact parameter $C$.
Indeed, we find (see the inset of Fig.\ \ref{resolved}) that the high-$n$ asymptote of the occupation probability $\sum_{mk} |\phi_{mkn}|^2 \sim 1/n^4$.
These high-$n$ states are just plane waves near the center of the trap where the polaron is localized, because at very high energies the details of the potential are irrelevant.
We thus obtain the characteristic decay $n_q \sim C/E_n^2 \sim C/q^4$, where $q$ is momentum.
For small values of $|U/J|$ we recover the weakly interacting limit $C \propto U^2$.

Our approach highlights a general problem with single-band models in the sense that effects related to the contact may be neglected in an uncontrolled way; the $1/q^4$-regime is entered when $r_0 \ll 1/q \ll \mathcal{L}$, where $r_0$ is the range of the inter-particle potential and $\mathcal{L}$ is any relevant length scale of the problem \cite{Zwerger2012a}.
In the results of Fig.\ \ref{resolved}, the contact regime is reached only for length scales shorter than the lattice spacing $k_\text{L}^{-1}$, for a wide range of interaction strengths $U/J$.
Therefore, caution should be applied when using single-band models even in the weakly interacting regime, as essential physics may be missed.

It would also be of interest to investigate the influence of $\bar{n}$ on the crossover to the contact regime.
In the limit where the gas is very dilute, i.e.\ $\bar{n} \ll 1$, the interparticle spacing will become very large relative to the lattice spacing.
Then it is plausible that the universal behavior will show up for $k < k_\text{L}$.
Unfortunately, it is not numerically feasible with our method to study the case of large $L$ with sufficient number of majority component particles, so that $\bar{n} \ll 1$.

\section{Conclusions}

In conclusion, we have performed a detailed analysis of a (spin down) impurity immersed in a sea of $N$ (spin up) fermions in a lattice with an additional harmonic trapping potential.
We find that the RSVA is accurate over a large range of the interaction strength $U$, from strongly attractive to moderately repulsive interactions.
Furthermore, we compute the contribution from higher bands in this system and find that the lowest band approximation breaks down even at relatively high values of the lattice depth $V_0/E_\text{R}$ if a sufficiently strong harmonic trapping potential is also present.
Finally, we derive a method to compute the Tan contact parameter using the RSVA in trapped highly polarized systems.
Our variational approach is general, simple and numerically efficient and can be readily generalized to an arbitrary external potential $V(x)$, higher dimensions and mass-imbalanced systems.

\section*{Acknowledgments}

We thank M.O.J. Heikkinen, J.-P. Martikainen and N.T. Zinner for insightful discussions. 
This work was supported by the Academy of Finland through its Centers of Excellence Programme (2012-2017) and under Projects Nos. 135000, 141039, 251748, 263347 and 272490. 
Computing resources were provided by CSC-the Finnish IT Centre for Science and LAPACK \cite{Lapack1999} was used for our computations.

\appendix
\section{Detailed derivation of the variational method}

To describe the impurity, we consider the following Hamiltonian:
\begin{equation}
 \mathcal{H} = \mathcal{H}_0 + \mathcal{H}_{\text{int}},
\end{equation}
where $\mathcal{H}_0$ is the single-particle Hamiltonian without inter-particle interactions (in the canonical ensemble).
This non-interacting Hamiltonian is given by:
\begin{align}
 \mathcal{H}_0 = & \int dx \sum_\sigma \psi_{\sigma}^\dagger(x) \Big[ -\hbarm \frac{\pa^2}{\pa x^2} + \nu x^2 + \nonumber \\
 & \frac{V_0}{2} (-\cos(2\pi k_\text{L} x) + 1) \Big] \psi_\sigma(x),
\end{align}
where $\psi_{\sigma}^{(\dagger)}(x)$ destroys (creates) a particle with spin $\sigma \in \{\uparrow,\downarrow\}$, $m$ is the mass of a particle (we assume no mass imbalance), $\nu$ measures the strength of the harmonic trapping potential, $V_0$ measures the depth of the periodic part of the potential and $k_\text{L}$ determines the periodicity of the lattice.
The interaction part of the Hamiltonian is (assuming contact interactions):
\begin{equation}
 \mathcal{H}_\text{int} = g \int dx \psiupdg \psidndg \psidn \psiup.
\end{equation}

We can calculate the eigenfunctions and eigenvalues of $\mathcal{H}_0$ numerically. 
Let us define the creation operators for the particles in the eigenstates of the Hamiltonian $\mathcal{H}_0$ as $a_{\sigma n}^\dagger$ where $\sigma$ is the spin index and $n = 0$ is the ground state. 
A possible technique for finding the approximate ground state for this Hamiltonian is a variational ansatz \cite{Chevy2006a}.
In the ansatz, one restricts the possible particle-hole excitations of the system to one, and neglects the probability of multiple particle-hole excitations.
In our basis of choice, it reads (note that in order to avoid double counting, it is necessary to demand that $m \neq k$):
\begin{equation}
 |\Psi \rangle = \sum_l \phi_l a_{\downarrow l}^\dagger  |0\rangle + \sum_{mkn} \phi_{mkn} a_{\uparrow m}^\dagger a_{\uparrow k} a_{\downarrow n}^\dagger |0 \rangle, \label{chevyansatz}
\end{equation}
where $|0\rangle$ represents the vacuum, that is, the majority component atoms filled up until the Fermi surface in the non-interacting ($g = 0$) state.
Since we are considering a fixed number $N$ of majority component particles, determining the state $|0\rangle$ is trivial -- it just consists of the lowest $N$ eigenstates.
The coefficients $\phi_l$ and $\phi_{mkn}$ are variational parameters, which are to be determined.
Note that in the case of zero inter-particle interactions ($g=0$), the solution is obtained by setting $\phi_0 = 1$, $\phi_l = 0$ for $l > 0$ and $\phi_{mkn} = 0$.

We are interested in the ground state energy of the impurity. 
For this purpose, let us now calculate the expectation value of the Hamiltonian $\langle \mathcal{H} \rangle = \langle \mathcal{H}_0 \rangle + \langle \mathcal{H}_{\text{int}} \rangle$. First consider the non-interacting part of the Hamiltonian:
\begin{align}
 &\langle \mathcal{H}_0 \rangle = \Big(\langle 0 | \sum_l \phi_l^* a_{\downarrow l} + \sum_{mkn} \langle 0 | \phi_{mkn}^* a_{\downarrow n} a_{\uparrow k}^\dagger a_{\uparrow m} \Big) H_0  \nonumber \\
 & \Big( \sum_l a_{\downarrow l}^\dagger \phi_l |0 \rangle + \sum_{mkn} \phi_{mkn} a_{\uparrow m}^\dagger a_{\uparrow k} a_{\downarrow n}^\dagger |0\rangle \Big) \nonumber \\
 & = \sum_l E_l |\phi_l|^2 + \sum_{mkn} \Delta E_{mkn} |\phi_{mkn}|^2.
\end{align}
Here the energies $E_l$ are the eigenvalues of the Hamiltonian $H_0$ ($l = 0$ is the ground state of the non-interacting system) and $\Delta E_{mkn} = (E_n - E_k + E_m)$.
It is straightforward to see that since $E_m > E_k$ (the operator $a_{\uparrow m}^\dagger$ cannot create particles in states that are already occupied) the ground state of the non-interacting system is the polaron in the lowest eigenstate of $\mathcal{H}_0$ with unit probability, as expected. 

Now consider the interaction part of the Hamiltonian. 
First we write it in the eigenfunction basis:
\begin{equation}
 \mathcal{H}_{\text{int}} = g \sum_{ijpq} U_{ijpq} a_{\uparrow i}^\dagger a_{\uparrow j} a_{\downarrow p}^\dagger a_{\downarrow q},
\end{equation}
where $U_{ijpq} = \int dx \alpha_i^*(x) \alpha_j(x) \beta_p^*(x) \beta_q(x)$.
Here the $\alpha$ and $\beta$ functions correspond to the eigenfunctions for the majority component and the impurity respectively, so that for instance $\psi_\uparrow^\dagger(x) = \sum_{i} \alpha^*_i(x) a_{\uparrow i}^\dagger$.
These eigenfunctions can be chosen to be the same, although it will turn out to be useful to choose a different basis for the impurity.
We remark that in our implementation the Hamiltonian is real and symmetric and both the eigenfunctions and the variational coefficients can be taken to be real.
However, for completeness' sake, we will continue to treat these quantities as complex.

It now follows that $\langle \mathcal{H}_\text{int} \rangle = \langle \mathcal{H}_\text{int} \rangle_1 + \langle \mathcal{H}_\text{int} \rangle_2 + \langle \mathcal{H}_\text{int} \rangle_3$, where the first term is a mean-field term and some careful bookkeeping is required when we consider the other terms. 
The first term is given by:
\begin{align}
 & \langle \mathcal{H}_\text{int} \rangle_1 = g \sum_{ll'ijpq} U_{ijpq} \phi_l^* \phi_{l'}  \langle a_{\downarrow l} a_{\uparrow i}^\dagger a_{\uparrow j} a_{\downarrow p}^\dagger a_{\downarrow q} a_{\downarrow l'}^\dagger  \rangle.
\end{align}
For the inner product to be non-zero (the states are obviously orthogonal) we must have $q = l'$, $i = j$ and $p = l$, so that:
\begin{align}
 \langle \mathcal{H}_\text{int} \rangle_1 & =  g \sum_{ill'} U_{iill'} \phi_l^* \phi_l' = g \sum_{ll'} U_{ll'} \phi_l^* \phi_{l'}, 
\end{align}
where $U_{ll'} = \int dx n_\uparrow(x) \beta_l(x) \beta_{l'}(x)$.
This is a Hartree energy-like term, where $n_\uparrow(x)$ is the majority component density.
The second term is given by:
\begin{align}
 & \langle \mathcal{H}_\text{int} \rangle_2 = \sum_{m'k'n'} \langle  \phi_{m'k'n'}^* a_{\downarrow n'} a_{\uparrow k'}^\dagger a_{\uparrow m'} \nonumber \\
 & \Big( g \sum_{ijpq} U_{ijpq} a_{\uparrow i}^\dagger a_{\uparrow j} a_{\downarrow p}^\dagger a_{\downarrow q} \Big) \sum_{mkn} \phi_{mkn} a_{\uparrow m}^\dagger a_{\uparrow k} a_{\downarrow n}^\dagger \rangle.
\end{align}
Analogous to the previous case, we have $n = q$ and $p = n'$:
\begin{align}
 & \langle \mathcal{H}_\text{int} \rangle_2 = \sum_{m'k'n'} \langle  \phi_{m'k'n'}^*  a_{\uparrow k'}^\dagger a_{\uparrow m'} \nonumber \\
 & \Big( g \sum_{ijn} U_{ijnn'} a_{\uparrow i}^\dagger a_{\uparrow j} \Big) \sum_{mk} \phi_{mkn} a_{\uparrow m}^\dagger a_{\uparrow k}  \rangle.
\end{align}
Now there are three ways to make sure the inner product is non-zero:
\begin{align}
 & k=k', m=m', i=j, \\
 & k=i, m=m', k'=j, \\
 & k=k', m=j, i=m'
\end{align}
so that (note the minus sign because of the change in the ordering of operators, i.e. Wick's theorem):
\begin{align}
 & \langle \mathcal{H}_\text{int} \rangle_2 =  g \sum_{imknn'} \phi_{mkn'}^* \phi_{mkn} U_{iinn'} \nonumber \\
 & - g \sum_{mkk'nn'} \phi_{mk'n'}^* \phi_{mkn} U_{kk'n'n} \nonumber \\
 & + g \sum_{mm'knn'} \phi_{m'kn'}^* \phi_{mkn} U_{m'mn'n}.
\end{align}
The occupation number probabilities $n_i$ for a certain state $i$ at zero temperature are now implicit in the summation. 
Writing them explicitly:
\begin{align}
 & \langle \mathcal{H}_\text{int} \rangle_2 = g \sum_{mknn'} \phi_{mkn'}^* \phi_{mkn} U_{nn'} n_k(1-n_m) \nonumber \\ 
 &  - g \sum_{mkk'nn'} \phi_{mk'n'}^* \phi_{mkn} U_{kk'n'n}  n_kn_{k'}(1-n_m) \nonumber \\
 & + g \sum_{mm'knn'} \phi_{m'kn'}^* \phi_{mkn} U_{m'mn'n}  n_k(1-n_m)(1-n_{m'}).
\end{align}
The third term is less complicated and is given by:
\begin{align}
 &\langle \mathcal{H}_\text{int} \rangle_3 = \sum_{mknlijpq} \langle \phi_{mkn}^* a_{\downarrow n} a_{\uparrow k}^\dagger a_{\uparrow m} \nonumber \\
 & \Big( g U_{ijpq} a_{\uparrow i}^\dagger a_{\uparrow j} a_{\downarrow p}^\dagger a_{\downarrow q} \Big) a_{\downarrow l}^\dagger \phi_l \rangle + \text{h.c.} \nonumber \\
 & = 2g \Re \Big[ \sum_{mknl} \phi_{mkn}^* \phi_l U_{mknl} n_k(1-n_m) \Big].
\end{align}
We can now determine the variational coefficients $\phi_l$ and $\phi_{mkn}$.
Consider the sum of terms $\langle \mathcal{H} \rangle = \langle \mathcal{H}_0 \rangle + \langle \mathcal{H}_\text{int} \rangle_1 + \langle \mathcal{H}_\text{int} \rangle_2 + \langle \mathcal{H}_\text{int} \rangle_3$.
It follows that $\frac{\pa}{\pa \phi_l^*} \langle \mathcal{H} \rangle = \frac{\pa}{\pa \phi_l^*} E\langle \Psi|\Psi \rangle = E \phi_l$.
The constant $E$ is a Lagrange multiplier, which can be identified with the energy of the impurity.
A similar procedure gives another set of equations for the coefficients $\phi_{mkn}$.
Following this recipe, we obtain:
\begin{align}
 & \frac{\pa E \langle \psi |\psi \rangle}{\pa \phi_l^*} =  E_l \phi_l + g\phi_l \sum_{l'} U_{ll'}  \nonumber \\
 & +g \sum_{mkn} \phi_{mkn} U_{kmln} n_k (1-n_m) = E\phi_l.
\end{align}
The differentiation with respect to $\phi_{mkn}^*$ gives:
\begin{align}
 & \frac{\pa E\langle \psi|\psi \rangle}{\pa \phi_{mkn}^*} = (E_n - E_k + E_m) \phi_{mkn}  \nonumber \\
 & + g \phi_l U_{kmln} n_k (1-n_m) \nonumber \\
 & + g\sum_{n'} \phi_{mkn'} U_{nn'} n_k(1-n_m) \nonumber \\
 & - g\sum_{k'n'} \phi_{mk'n'} U_{k'knn'} n_{k'}n_k(1-n_m) \nonumber \\
  & + g\sum_{m'n'} \phi_{m'kn'} U_{mm'nn'} n_k(1-n_m)(1-n_{m'}).
\end{align}
Let us introduce a new function to shorten notations:
\begin{align}
 & \Gamma_{mkn} = \sum_{n'} \phi_{mkn'} U_{nn'} n_k(1-n_m) \nonumber \\
 & - \sum_{k'n'} \phi_{mk'n'} U_{k'knn'}  n_{k'}n_k(1-n_m) \nonumber \\
  & + \sum_{m'n'} \phi_{m'kn'} U_{mm'nn'} n_k(1-n_m)(1-n_{m'}). \label{gamma}
\end{align}
Now we can write:
\begin{align}
 & E \phi_l = E_l \phi_l + g \sum_{l'} \phi_{l'} U_{ll'} + g \sum_{mkn} \phi_{mkn} U_{mknl}, \label{energyiter1} \\
 & E \phi_{mkn} = \Delta E_{mkn} \phi_{mkn} + g \Gamma_{mkn} + g \sum_l \phi_l U_{mknl}.
\end{align}
This system of equations can be solved iteratively starting e.g.\ from an initial state where $\phi_0 = 1$ and all other coefficients are zero, i.e.\ the ground state of the non-interacting system.

The problem with the above scheme is that the fixed-point iteration is not guaranteed to converge, especially if the initial state is far from the ground state.
We may find a metastable state, or worse, the iteration might not converge at all.
A convenient alternative choice (although not necessarily the best) is writing the polaron eigenstates in terms of a ``mean-field'' basis. 
We consider the slightly different Hamiltonian to describe the polaron:
\begin{align}
 & \mathcal{H}_\text{0,MF} =  \int dx \psi_\downarrow^\dagger(x) \Big[  -\hbarm \frac{\pa^2}{\pa x^2} + \frac{1}{2} m\omega^2 x^2 + \nonumber \\
 & \frac{V_0}{2} (-\cos(2\pi k_\text{L} x) + 1) + gn_\uparrow(x) \Big] \psi_\downarrow(x),
\end{align}
where $n_\uparrow(x)$ is the majority component density in the non-interacting system.
The eigenfunctions of this Hamiltonian can still be obtained through an eigenvalue solver; one just needs to calculate the eigenfunctions of the non-interacting Hamiltonian first, compute $n_\uparrow(x)$ from the resulting eigenbasis, and use the result to compute the eigenfunctions in the new basis for the polaron.
Since the total Hamiltonian describing the system is unchanged, we subtract the term that was added from the interaction Hamiltonian, so that:
\begin{align}
 &\mathcal{H} = \mathcal{H}_\text{0,MF} + \mathcal{H}_\text{int} + \mathcal{H}_\text{MF}, \,\,\,\, \text{where} \nonumber \\
 & \mathcal{H}_\text{MF} = -g \int dx \psi_\downarrow^\dagger(x) \psi_\downarrow(x) n_\uparrow(x),
\end{align}
and calculate the expectation value:
\begin{align}
 & \langle  \mathcal{H}_\text{MF} \rangle =  -g \langle  \Big( \sum_l \phi_l^* a_{\downarrow l} + \sum_{mkn} \phi_{mkn}^* a_{\downarrow n} a_{\uparrow k}^\dagger a_{\uparrow m} \Big) \nonumber \\
  & \sum_{ij} U_{ij}  a_{\downarrow i}^\dagger a_{\downarrow j} \Big(\sum_{l'} a_{\downarrow l}^\dagger \phi_{l'} + \sum_{m'k'n'} \phi_{m'k'n'}^* a_{\uparrow m'}^\dagger a_{\uparrow k'}  a_{\downarrow n'}^\dagger \Big)  \rangle.
\end{align}
We have four terms. The first one is simple:
\begin{align}
 & \langle  \sum_{ll'ij} U_{ij} \phi_l^* a_{\downarrow l}   a_{\downarrow i}^\dagger a_{\downarrow j} a_{\downarrow l'}^\dagger \phi_{l'}  \rangle = -g \sum_{ll'} \phi_l \phi_{l'} U_{ll'},
\end{align}
which precisely cancels the $U_{ll'}$-term in eq.\ (\ref{energyiter1}).
The second and third terms (the cross-terms) drop out because of the requirement that $m \neq k$.
The final term is given by:
\begin{align}
 & -g\langle  \sum_{mkn} \phi_{mkn}^* a_{\downarrow n} a_{\uparrow k}^\dagger a_{\uparrow m} \sum_{ij} U_{ij} a_{\downarrow i}^\dagger a_{\downarrow j} \nonumber \\
 & \sum_{m'k'n'} \phi_{m'k'n'}^* a_{\uparrow m'}^\dagger a_{\uparrow k'}  a_{\downarrow n'}^\dagger \rangle \nonumber \\
 & = -g\sum_{mknn'} \phi_{mkn}^* \phi_{mkn'} n_k (1-n_m) U_{nn'},
\end{align}
which cancels the term containing the density in $\Gamma_{mkn}$ -- the first term in eq.\ (\ref{gamma}).
The change of basis thus conveniently simplifies the variational calculation by removing two of the terms.

\bibliographystyle{apsrev4-1}
\bibliography{ref}

\end{document}